\documentclass[12pt]{article}
\begin{document}
\title{Julian Schwinger\\(1918-1994)}

\author{K. A. Milton\\Homer L. Dodge Department of Physics and Astronomy,\\
University of Oklahoma, Norman, OK 73019}
\maketitle

Julian Schwinger's influence on Twentieth Century science is profound
and pervasive.  Of course, he is most famous for his renormalization
theory of quantum electrodynamics, for which he shared the Nobel Prize
with Richard Feynman and Sin-itiro Tomonaga.  But although this triumph
was undoubtedly his most heroic accomplishment, 
his legacy lives on chiefly through
subtle and elegant work in classical electrodynamics, quantum variational
principles, proper-time methods, quantum anomalies, dynamical mass generation,
partial symmetry, and more.  Starting as just a boy, he rapidly became the
pre-eminent nuclear physicist in the late 1930s, led the theoretical
development of radar technology at MIT during World War II, and then, soon
after the war, conquered quantum electrodynamics, and became the leading
quantum field theorist for two decades, before taking a more iconoclastic
route during his last quarter century.

Given his commanding stature in theoretical physics for decades
it may seem puzzling why he is relatively unknown now to the educated public,
even to many younger physicists, while Feynman is a cult figure with his
photograph needing no more introduction than Einstein's.  This relative
obscurity is even more remarkable,
in view of the enormous number of eminent physicists, as well as other leaders
in science and industry, who received their Ph.D.'s under Schwinger's
direction, while Feynman had but few.  In part, the answer lies in Schwinger's
retiring nature and reserved demeanor.  Science, research and teaching, were
his life, and he detested the limelight.  Generally, he was not close to
his students, so few knew him well.  He was a gracious host and a good
conversationalist, and had a broad knowledge of many subjects, but he was
never one to initiate a relationship or flaunt his erudition.
His style of doing physics was also difficult to penetrate.  Oppenheimer
once said that most people gave talks to show others how to do the calculation,
while Schwinger gave talks to show that only {\em he\/} could do it.
Although a commonly shared view, this witticism is unkind and untrue.
He was, in fact, a superb teacher, and generations of physicists, students
and faculty alike, learned physics at his feet. On the one hand he was
a formalist, inventing formalisms so powerful that they could lead, at
least in his hands, unerringly to the correct answer. He did not, therefore,
display the intuitive visualizations, for example, that Feynman commanded,
which eventually took over the popular and scientific culture.  But,
more profoundly, he was a phenomenologist.  Ironically, even some of his
own students criticized him in his later years for his phenomenological
orientation, not recognizing that he had, from his earliest experiences
in nuclear physics, insisted in grounding theoretical physics in the
phenomena and data of experiment.  Isidor I. Rabi, who discovered Schwinger
and brought him to Columbia University, generally had a poor opinion
of theoretical physicists.  But Rabi was always very impressed with
Schwinger because in nearly every paper, he `got the numbers out' to
compare with experiment.  Even in his most elaborate field-theoretic
papers he was always concerned with making contact with the real
world, be it quantum electrodynamics, or strongly interacting hadrons.

His strong phenomenological bent eventually led him away from the mainstream
of physics.  Although he had given the basis for what is now called the
Standard Model of elementary particles in 1957, he never could accept
the existence of quarks because they had no independent existence outside
of hadrons.  He came to appreciate the notion of supersymmetry,
but he rejected notions of `Grand Unification' and of `Superstrings'
not because of their structure but because he saw them as preposterous
speculations, based on the notion that nothing new remains to be found
from 1 TeV to $10^{19}$ GeV.  He was sure that totally new, unexpected
phenomena were waiting just around the corner.  This seems a reasonable
view, but it resulted in a self-imposed isolation, in contrast, again, to
Feynman, who contributed mightly to the theory of partons and quantum
chromodynamics up to the end.

\section*{A Brief Life of Schwinger}
A full biography of Julian Schwinger has been published \cite{MM},
as well as a selection of his most important papers \cite{QL}.
Here we will sketch a brief outline of Schwinger's life and work, referring
the interested reader to the biography for more details.  An excellent
100-page account of Schwinger's career through 1950 may also be found in
Schweber's history of quantum electrodynamics \cite{SSS}.

Julian Schwinger was born in Manhattan, New York City, on February 12, 1918,
to rather well-off middle-class parents.  His father was a well-known
designer of women's clothes.  He had a brother Harold seven years older than
himself, whom Julian idolized as child.  Harold claimed that he taught
Julian physics until he was 13. Although Julian was recognized
as intelligent in school, everyone thought Harold was the bright one.
(Harold in fact eventually became a well-known lawyer, and his mother
always considered him as the successful son, even after Julian received the
Nobel Prize.)  The Depression cost Julian's father his business, but he
was sufficiently appreciated that he was offered employment by other
designers; so the family survived, but not so comfortably as before.
It did mean that Julian would have to rely on free education, which New York
well-provided in those days: A year or two at Townsend Harris High School,
a public preparatory school feeding into City College, where Julian
matriculated in 1933.  Julian had already discovered physics, first through
Harold's {\it Encyclopedia Britannica\/} at home, and then through the
remarkable institution of the New York Public Library.  At City College
Julian was reading and digesting the latest papers from Europe, and starting
to write papers with instructors who were, at the same time, graduate
students at Columbia and NYU.  He no longer had the time to spend in the
classroom attending lectures.  In physics and mathematics he was able
to skim the texts and work out the problems from first principles, 
frequently leaving the professors baffled with his original, unorthodox
solutions, but it was not so simple in history, English, and German.
City College had an enormous number of required courses then in all subjects.
His grades were not good, and he would have flunked out if the College had
not also had a rather forgiving policy toward grades.

Not only was Julian already reading the literature at City College, but
he quickly started to do original research.
Thus before he left the City College, Schwinger wrote a  paper entitled
`On the
 Interaction of Several Electrons,'
in which he introduced a procedure that
he would later call the interaction representation to 
describe the scattering of
spin-1/2 Dirac particles, electron-electron scattering or M\o ller 
scattering.  This paper he wrote entirely on his own, but showed it
to no one, nor did he submit it to a journal.  It was `a little practice
in writing,' but it was a sign of great things to come.

It was Lloyd Motz, one the instructors at City College, 
who had heard about Julian from Harold, and with whom Julian
 was writing  papers,
who introduced him to Rabi.  Then, in a conversation between Rabi and Motz
over the Einstein, Rosen, Podolsky
paper \cite{erp}, which had just appeared, Julian's
voice appeared with the resolution of a difficulty through the completeness
principle, and Schwinger's career was assured.  Rabi, not without some
difficulty, had Schwinger transferred to Columbia with a scholarship, 
and by 1937 he had
7 papers published, which constituted his Ph.D. thesis, even though his
bachelor's degree had not yet been granted.  The papers which Julian 
wrote at Columbia were on both
theoretical and experimental physics, and Rabi prized Julian's ability
to obtain the numbers  to compare with experiment.
 The formal awarding of the
Ph.D. had to wait till 1939 to satisfy a University regulation.  In
the meantime, Schwinger was busy writing papers (one, for example,
laid the foundation for the theory of nuclear magnetic resonance),
and spent a somewhat lonely, but productive winter of 1937
 in Wisconsin, where he provided the groundwork for his
prediction that the deuteron had an electric quadrupole moment,
 independently
confirmed by his experimental colleagues at Columbia a year later \cite{quad},
both announced at the Chicago meeting of the American Physical Society in
November 1938.
Wisconsin confirmed his predilection for working at night, so as not
to be `overwhelmed' by his hosts, Eugene Wigner and Gregory Breit.

By 1939, Rabi felt Schwinger had outgrown Columbia, so with a NRC Fellowship,
he was sent to J. Robert 
Oppenheimer in Berkeley.  This exposed him to new fields: quantum 
electrodynamics (although as we recall, 
he had written an early, unpublished paper on the
subject while just 16) and cosmic-ray physics, but he mostly continued to
work on nuclear physics.  He had a number of collaborations; the most
remarkable was with William Rarita, who was on sabbatical from Brooklyn
College:  Rarita was Schwinger's `calculating arm' on a series of papers
extending the notion of nuclear tensor forces which he had conceived in
Wisconsin over a year earlier.  Rarita and Schwinger also wrote the
prescient paper on spin-3/2 particles, which was to be influential decades
later with the birth of supergravity.

The year of the NRC Fellowship was followed by a second year at Berkeley
as Oppenheimer's assistant.  They had already written
an important paper together which
would prove crucial several years later:  Although Oppenheimer was
happy to imagine new interactions, Schwinger showed that an anomaly in
fluorine decay could be explained by the existence of vacuum polarization,
that is, by the virtual creation of electron-positron pairs.  This gave
Schwinger a head start over Feynman, who for years suspected that vacuum
polarization did not exist.

After two years at Berkeley, Oppenheimer and Rabi arranged a real job for
Schwinger: He became first an instructor, then an Assistant Professor
at Purdue University, which had acquired a number of bright young
physicists under the leadership of Karl Lark-Horowitz.  But the war was
impinging on everyone's lives, and Schwinger was soon recruited into
the work on radar.  The move to the MIT Radiation Laboratory took place in
1943.  There Schwinger rapidly became the theoretical leader, even though
he was seldom seen, going home in the morning just as others were arriving.
He developed powerful variational methods for dealing with complicated
microwave circuits, expressing results in terms of quantities the engineers
could understand, such as impedance and admittance.  These methods, and
the discoveries he made there concerning the reality of the electromagnetic
mass, would be invaluable for his work on quantum electrodynamics a few
years later.  As the war wound down, physicists started thinking about
 new accelerators, since the pre-war cyclotrons had been defeated by
relativity, and Schwinger became a leader in this development: he proposed
a microtron, a accelerator based on acceleration through microwave cavities,
developed the theory of stability of synchrotron orbits, and most importantly,
worked out in detail the theory of synchrotron radiation, at a time when
many thought that such radiation would be negligible because of destructive
interference.\footnote{Material based on his Radiation Lab work has now
been published \cite{ER}.}

Although he never really published his considerations on self-reaction,
he viewed that understanding as the most important part of his work on
synchrotron radiation:
`It was a useful thing for me for what was
to come later in electrodynamics, because the technique I used for calculating
the electron's classical radiation was one of self-reaction, and I did it
relativistically, and it was a situation in which I had to take seriously the 
part
of the self-reaction which was radiation, so why not take seriously the part of
the self-reaction that is mass change?  In other words, the ideas of mass
renormalization and relativistically handling them were already present at this
classical level.' \cite{js}

At first it may seem strange that Schwinger, by 1943 the leading nuclear
theorist, should not have gone to Los Alamos, where nearly all his
colleagues eventually settled for the duration.
There seem to be at least three reasons why Schwinger stayed at the
Radiation Laboratory throughout the war.
\begin{itemize}
\item The reason he most often cited later in life was one of moral
repugnance.  When he realized the destructive power of what was being
constructed at Los Alamos, he wanted no part of it.  In contrast,
the radiation lab was developing a primarily defensive technology, radar,
which had already saved Britain.
\item He believed that the problems to solve at the Radiation Laboratory
were more interesting.  Both laboratories were involved 
largely in
engineering, yet although Maxwell's equations were certainly well known,
the process of applying them to waveguides required the development of
special techniques that would prove invaluable to Schwinger's later
career.
\item Another factor probably was Schwinger's fear of being overwhelmed.
In Cambridge he could live his own life, working at night when no one
was around the lab.  This privacy would have been much more difficult
to maintain in the microworld of Los Alamos.
Similarly, the working conditions at the Rad Lab were much freer
than those at Los Alamos.  Schwinger never was comfortable in a team
setting, as witness his later aversion to the atmosphere at the
Institute for Advanced Study.
\end{itemize}

In 1945 Harvard offered Schwinger an Associate Professorship, which he
promptly accepted, partly because in the meantime he had met his future
wife Clarice Carrol.  Counteroffers quickly appeared, from Columbia,
Berkeley, and elsewhere, and Harvard shortly made Schwinger the youngest
full professor on the faculty to that date. There Schwinger quickly established
a pattern that was to persist for many years---he taught brilliant courses
on classical electrodynamics, nuclear physics, and quantum mechanics,
surrounded himself with a devoted coterie of graduate students and
post-doctoral assistants, and conducted incisive research that set the
tone for theoretical physics throughout the world.  Work on classical
diffraction theory, begun at the Radiation Lab, continued for several years
largely due to the presence of Harold Levine, whom Schwinger had brought along
as an assistant.  Variational methods, perfected in the electrodynamic
waveguide context, were rapidly applied to problems in nuclear physics.
But these were old problems, and it was quantum electrodynamics that
was to define Schwinger's career.

\section*{Quantum Electrodynamics}

But it took new experimental data to catalyze this development.  That
data was presented at the famous Shelter Island meeting held in June 1947,
a week before Schwinger's wedding.  There he, Feynman, Victor Weisskopf,
Hans Bethe, and the other
participants learned the details of the new experiments of Lamb and
Retherford \cite{lamb}
that confirmed the pre-war Pasternack effect, showing a splitting between
the $2S_{1/2}$ and $2P_{1/2}$ states of hydrogen, that should be degenerate according
to Dirac's theory.  In fact, on the way to the conference, Weisskopf and
Schwinger speculated that quantum electrodynamics could explain this effect,
and outlined the idea to Bethe there, who worked out the details, 
non-relativistically, on his famous train ride to Schenectady 
after the meeting \cite{bethe}.
But the other experiment announced there was unexpected: This was the
experiment by Rabi's group and others \cite{g-2} of the hyperfine anomaly
 that would prove to
mark the existence of an anomalous magnetic moment of the electron,
\begin{equation}
\mbox{\boldmath{$\mu$}}=g{e\over2m}{\bf S},
\end{equation}
 differing
from the value $g=2$ again predicted by Dirac.  Schwinger immediately
saw this as the crucial calculation to carry out first, because it
was purely relativistic, and much cleaner to understand theoretically,
not involving the complication of bound states.  However, he was delayed
three months in beginning the calculation because of an extended honeymoon
through the West.  He did return to it in October, and by December 1947
had obtained the result $g/2=1+{\alpha/2\pi}$, completely consistent with
experiment.  He also saw how to compute the relativistic Lamb shift (although
he did not have the details quite right), and found the error in the
pre-war Dancoff calculation of the radiative correction to electron scattering
by a Coulomb field \cite{dancoff}. In effect, he had solved all the fundamental
problems that had plagued quantum electrodynamics in the 1930s:  The
infinities were entirely isolated in quantities that renormalized the
mass and charge of the electron.  Further progress, by himself and others,
was thus a matter of technique.

\section*{Covariant Quantum Electrodynamics}
During the next two years Schwinger developed two new approaches to quantum 
electrodynamics.
His original approach, which made use of successive canonical transformations
to isolate the infinities, while
sufficient for calculating the anomalous magnetic moment of the electron, was
noncovariant, and as such, led to inconsistent results.  In particular, 
the magnetic moment appeared
also as part of the Lamb shift calculation, through the coupling with the 
electric field implied by
relativistic covariance; but the noncovariant scheme gave the 
wrong coefficient. (If the coefficient
were modified by hand to the right number, what turned out to be the correct 
relativistic value for
the Lamb shift emerged, but what that was was unknown in January 1948,
when Schwinger announced his results at the APS meeting in New York.)  
So first at the Pocono Conference in
April 1948,  then in the Michigan Summer School that year, and finally in a 
series of three monumental
papers,  `Quantum Electrodynamics I, II, and III,' 
 Schwinger set forth his covariant approach to QED.  At about
the same time Feynman was formulating his covariant path-integral approach; 
and although Feynman's
presentation at Pocono was not well-received, Feynman and Schwinger compared notes and realized
that they had climbed the same mountain by different routes.  
Feynman's systematic papers \cite{feynman} were
published only after Dyson \cite{dyson} 
had proved the equivalence of Schwinger's and Feynman's schemes.

It is worth remarking that Schwinger's approach was conservative.  He took 
field theory at face value,
and followed the conventional path of Pauli, Heisenberg, and Dirac \cite{phd}.
  His genius was to recognize that
the well-known divergences of the theory, which had stymied all pre-war 
progress, could be
consistently isolated in renormalization of  charge and mass.  This bore a 
superficial resemblance
to the ideas of Kramers advocated as early as 1938 \cite{kramers},
 but Kramers proceeded classically.  He had
insisted that first the classical theory had to be rendered finite and then 
quantized.  That idea was
a blind alley.  Renormalization of quantum field theory is unquestionably the 
discovery of Schwinger.
Feynman was more interested in finding an alternative to field theory, 
eliminating entirely the 
photon field in favor of action at a distance.  He was, by 1950, quite 
disappointed to realize that
his approach was entirely equivalent to the conventional electrodynamics, in 
which electron and photon fields are treated on the same footing.

As early as January 1948, when Schwinger was expounding his 
noncovariant QED to overflow crowds
at the American Physical Society meeting at Columbia University, he learned 
from Oppenheimer
of the existence of the work of Tomonaga carried out in Tokyo during the 
terrible conditions of wartime \cite{tomonaga}.
 Tomonaga had independently invented the `Interaction Representation'
  which Schwinger
had used in his unpublished 1934 paper, and had come up with a 
covariant version of the Schr\"odinger
equation as had Schwinger, which upon its Western rediscovery was dubbed
by Oppenheimer the Tomonaga-Schwinger
equation.  Both Schwinger and Tomonaga independently wrote the same equation,
a generalization of the Schr\"odinger equation to an arbitrary spacelike
surface $\sigma$, using nearly the same notation:
\begin{eqnarray}
i\hbar c {\delta\Psi[\sigma]\over\delta\sigma(x)}={\cal H}(x)\Psi[\sigma],
\label{tomeqn}
\end{eqnarray}
where $\cal H$ is the interaction Hamiltonian,
\begin{eqnarray}
{\cal H}(x)=-{1\over c}j_\mu(x)A_\mu(x),
\label{intham}
\end{eqnarray}
$j_\mu$ being the electric current density of the electrons, and $A_\mu$ the
electromagnetic vector potential.
 The formalism found by Tomonaga and his school was essentially identical to 
that developed
by Schwinger five years later; yet they at the time calculated nothing, 
nor did they discover 
renormalization.  That was certainly no reflection on the ability of the 
Japanese; Schwinger could not
have carried the formalism to its logical conclusion without the impetus of 
the postwar experiments,
which overcame prewar paralysis by showing that the quantum corrections
`were neither infinite nor zero, but finite and small, and 
demanded understanding.' \cite{RTQED}

However, at first Schwinger's covariant
calculation of the Lamb shift contained another error, the same as
Feynman's \cite{rpf}.
 `By this time I had forgotten the number
I had gotten by just artificially changing the wrong spin-orbit coupling.
Because I was now thoroughly involved with the covariant calculation and it
was the covariant calculation that betrayed me, because something went wrong
there as well.  That was a human error of stupidity.' \cite{js}  French and 
Weisskopf \cite{fw}
had gotten the right answer, `because they
put in the correct value of the magnetic moment and used it all the way 
through.  I, at an earlier stage, had done that, in effect, and also got the
same answer.' \cite{js}
  But now he and Feynman `fell into the same trap.  We were
connecting a relativistic calculation of high energy effects with a
nonrelativistic calculation of low energy effects, a la Bethe.'  Based
on the result Schwinger had presented at the APS meeting in January 1948,
Schwinger claimed priority for the Lamb shift calculation: `I had the
answer in December of 1947.  If you look at those [other] papers you will
find that on the critical issue of the spin-orbit coupling, they appeal
to the magnetic moment.  The deficiency in the calculation I did [in 1947]
was [that it was] a non-covariant calculation.  French and Weisskopf
were certainly doing a non-covariant calculation.  Willis 
Lamb \cite{kl} was doing
a non-covariant calculation.  They could not possibly have avoided these
same problems.' 
 The error Feynman and Schwinger
made had to do with the infrared problem that occurred in the relativistic
calculation, which was handled by giving the photon a fictitious mass.
`Nobody thought that if you give the photon a finite mass it will also affect
the low energy problem.  There are no longer the two transverse degrees
of freedom of a massless photon, there's also a longitudinal degree of
freedom.  I suddenly realized this absolutely stupid error, that a photon
of finite mass is a spin one particle, not a helicity one 
particle.'  Feynman \cite{feynman} 
was more forthright and apologetic in acknowledging his
error which substantially delayed the publication of the French and Weisskopf
paper, in part because he, unlike Schwinger, had published
his incorrect result \cite{rpf}.

\section*{Quantum Action Principle}
Schwinger learned from his competitors, particularly Feynman and Dyson. 
 Just as Feynman had
borrowed the idea from Schwinger
that henceforward would go by the name of Feynman parameters,
Schwinger recognized that the systematic approach of Dyson-Feynman was 
superior in higher orders.  So by 1949 he replaced the 
Tomonaga-Schwinger approach by a much more powerful engine,
the quantum action principle.  This was a logical outgrowth of the formulation 
of Dirac \cite{lagrangian}, as was
Feynman's path integrals; the latter was an integral approach, Schwinger's 
a differential.  The
formal solution of Schwinger's differential equations was Feynman's functional 
integral; yet while the
latter was ill-defined, the former could be given a precise meaning, and for 
example, required the
introduction of fermionic variables, which initially gave Feynman some 
difficulty. It may be fair to say, at the beginning of the new millennium,  
that while the path integral formulation of quantum field 
theory receives all the press, the most precise exegesis of field theory 
is provided by the functional differential
equations of Schwinger resulting from his action principle.

He began in the `Theory of Quantized Fields I'  by introducing 
a complete set of eigenvectors `specified by a spacelike 
surface $\sigma$ and the eigenvalues $\zeta'$ of a complete set of 
commuting operators constructed from field quantities attached to that
surface.'  The question is how to compute the transformation function
from one spacelike surface to another, that is, $(\zeta'_1,\sigma_1|\zeta''_2,
\sigma_2)$.  After remarking that this development, time-evolution,
must be described by a unitary transformation, he {\it assumed\/} that
any infinitesimal change in the transformation function must be given
in terms of the infinitesimal change in a quantum action operator, $W_{12}$,
or of a quantum Lagrange function $\cal L$.  This is the quantum dynamical
principle:
\begin{eqnarray}
&&\delta(\zeta'_1,\sigma_1|\zeta''_2,\sigma_2)={i\over\hbar}(\zeta'_1,\sigma_1|
\delta W_{12}|\zeta''_2,\sigma_2)\nonumber\\
&&\quad={i\over \hbar}(\zeta'_1,\sigma_1|\delta\int_{\sigma_2}^{\sigma_1}
(dx){\cal L}(x)|\zeta_2'',\sigma_2).
\label{qap}
\end{eqnarray}
Here, $\cal L$ is a relativistically invariant Hermitian function of
the fields and their derivatives,
\begin{eqnarray}
{\cal L}(x)={\cal L}(\phi^a(x),\partial_\mu\phi^a(x)),
\end{eqnarray}
where $a$ labels the different field operators of the system.
If the parameters of the system are not altered, the only changes
arise from those of the initial and final states, which changes are
effected by infinitesimal generating operators $F(\sigma_1)$, $F(\sigma_2)$,
expressed in terms of operators associated with the surfaces $\sigma_1$
and $\sigma_2$.  In this way, Schwinger deduced the {\it Principle of
Stationary Action}, 
\begin{eqnarray}
\delta W_{12}=F(\sigma_1)-F(\sigma_2),
\end{eqnarray}
from which the field equations may be deduced.  A series of six papers followed
with the same title, and the most important `Green's Functions of
Quantized Fields,' published in the Proceedings of the National Academy
of Sciences.

The paper `On Gauge Invariance and Vacuum Polarization,' 
submitted by Schwin\-ger
 to the {\it Physical Review\/} near the end of December 1950,
is nearly universally acclaimed as his greatest publication.  As his lectures
have rightfully been compared to the works of Mozart, so this might be
compared to a mighty construction of Beethoven, the 3rd Symphony, the {\it
Eroica}, perhaps.  It is most remarkable because it stands in splendid
isolation.  It was written over a year 
after the last of his series of papers on
his second, covariant, formulation of quantum electrodynamics
was completed: `Quantum Electrodynamics III. The Electromagnetic Properties
of the Electron---Radiative Corrections to Scattering' was submitted
in May 1949.  And barely two months later, in March 1951,
 Schwinger would submit the first
of the series on his third reformulation of quantum 
field theory, that
based on the quantum action principle, namely, 
`The Theory of Quantized Fields I.'  But 
`Gauge Invariance and Vacuum Polarization'  stands on its own,
and has endued the rapid changes in tastes and developments in quantum
field theory, while the papers in the other series are mostly of historical
interest now. Among many other remarkable developments, Schwinger discovered
here the axial-vector anomaly, nearly twenty years before its rediscovery and
naming by Adler, Bell, and Jackiw \cite{abj}.
 As Lowell Brown \cite{brown}
pointed out, `Gauge Invariance and Vacuum Polarization'
still has over one hundred citations per year, and is far and away Schwinger's 
most cited paper.\footnote{In 2005 the {\it Science Citation Index} lists
104 citations, out of a total of 458 citations to all of Schwinger's 
work.  These numbers have remained remarkably constant over ten years.}

So it was no surprise that in the late 1940s and early 1950s Harvard was the 
center of the world, as
far as theoretical physics was concerned.  Everyone, students and professors 
alike, flocked to
Schwinger's lectures.  Everything was revealed, long before publication; 
and not infrequently
others received the credit because of   Schwinger's reluctance to publish 
before the subject was ripe.
A case in point is the so-called Bethe-Salpeter equation \cite{bs},
 which as Gell-Mann and Low noted \cite{gml}, first
appeared in Schwinger's lectures at Harvard.  
At any one time, Schwinger had ten or twelve Ph.D.
students, who typically saw him but rarely.  In part, this was because he was 
available to see his
large flock but one afternoon a week, but most saw him only when absolutely 
necessary, because
they recognized that his time was too valuable to be wasted on trivial 
matters.  A student may have
seen him only a handful of times in his graduate career, but that was all the 
student required.
When admitted to his sanctum, students were never rushed, were listened to 
with respect, treated with kindness,
and given inspiration and practical advice.  One must remember that 
the student's  problems
were typically quite unrelated to what Schwinger himself was working on at 
the time; yet in a few
moments, he could come up with amazing insights that would keep the student 
going for weeks,
if not months.  A few students got to know Schwinger fairly well, and were 
invited to the Schwingers'
house occasionally; but most saw Schwinger primarily as a virtuoso in the 
lecture hall, and now and
then in his office.  A few faculty members were a bit more intimate, but 
essentially Schwinger was a very private person.

\section*{Field Theory}
Feynman left the field of quantum electrodynamics in 1950, regarding it as 
essentially complete.
Schwinger never did.  During the next fifteen years, he continued to explore 
quantum field theory,
trying to make it reveal the secrets of the weak and strong interactions.  
And he accomplished much.
In studying the relativistic structure of the theory, he recognized that 
all the physically significant
representations of the Lorentz group were those that could be derived from the 
`attached' four-dimensional
Euclidean group, which is obtained by letting the time coordinate become 
imaginary.  This idea was
originally ridiculed by Pauli, but it was to prove a most fruitful suggestion.
Related to this was the
CPT theorem, first given a proof for interacting systems by Schwinger in his 
`Quantized Field' papers
of the early 1950s, and elaborated later in the decade. 
 By the end of the 1950s, Schwinger, with his
former student Paul Martin, was applying field theory methods to many-body 
systems, which led to a
revolution in that field, and independently developed techniques which
opened up non-equilibrium statistical mechanics.
 Along the way, in what he considered rather modest papers, he discovered
Schwinger terms, anomalies in the commutation relations between 
field operators, and the Schwinger
model, still the only known example of dynamical mass generation.  
The beginning of a quantum field theory for non-Abelian fields was made; 
the original example of a non-Abelian field being that of
the gravitational field, he laid the groundwork for later canonical 
formulations of gravity. (See also \cite{adm}.)  Fundamental here were his
consistency conditions for a relativistic quantum field theory.

\section*{Measurement Algebra}
In 1950 or so, as we mentioned, Schwinger developed his action principle, 
which applies to any
quantum system, including nonrelativistic quantum mechanics.  Two years later, 
he reformulated
quantum kinematics, introducing symbols that abstracted the essential elements 
of realistic measurements.
This was measurement algebra, which yielded conventional Dirac 
quantum mechanics.  But although
the result was as expected, Schwinger saw the approach as of great value 
pedagogically, and as
providing a interpretation of quantum mechanics that was self-consistent.  He 
taught quantum mechanics
this way for many years, starting in 1952 at the Les Houches summer school; 
but only in 1959 did he
start writing a series of papers expounding the method to the world.  
He always intended to write a
definitive textbook on the subject, but only an incomplete version based 
on the 
Les Houches lectures ever appeared.  (In the last few years, Englert
brought his UCLA quantum mechanics
 lectures to a wider audience \cite{englertbook}.)

One cannot conclude a retrospective of Schwinger's work without mentioning
two other magnificent achievements in the quantum mechanical domain.
 He presented in 1952 a definitive development
of angular momentum theory derived in terms of oscillator variables in
 `On Angular Momentum,'  which was never properly
published; and he developed a `time-cycle' method of calculating
matrix elements without having to find all the wavefunctions in his
beautiful  `Brownian Motion of a Quantum Oscillator' (1961).  
We should also mention the famous Lippmann-Schwinger paper (1950),
which is chiefly remembered for what Schwinger considered a standard
exposition of quantum scattering theory, not for the variational methods
expounded there.

\section*{Electroweak Synthesis}
In spite of his awesome ability to make formalism work for him, Schwinger was 
at heart a
phenomenologist.  He was active in the search for higher symmetry; while he 
came up with $W_3$,
Gell-Mann found the correct approximate symmetry of hadronic states, $SU(3)$.  
Schwinger's
greatest success in this period was contained in his masterpiece, 
his 1957 paper `A Theory of
the Fundamental Interactions.'  Along with many other insights, 
such as the existence of two neutrinos and the $V-A$ structure of
weak interactions, Schwinger there laid the
groundwork for the electroweak unification.  He introduced two charged 
intermediate vector
bosons as partners to the photon, which couple to charged weak currents.  
That coupling is exactly that found in the standard model. A few years later,
his former student, Sheldon Glashow, as an outgrowth of his thesis, would 
introduce a neutral
heavy boson to close  the system to the modern $SU(2)\times  U(1)$ symmetry 
group \cite{glashow}; Steven
Weinberg \cite{weinberg} would complete the picture by 
generating the masses for 
the heavy bosons by
spontaneous symmetry breaking.  Schwinger did not have the details 
right in 1957, in 
particular because experiment seemed to disfavor the $V-A$ theory his 
approach implied,
but there is no doubt that Schwinger must be counted as the grandfather of the 
Standard Model on the basis on this paper.

\section*{The Nobel Prize and Reaction}
Recognition of Schwinger's enormous contributions had come early.  
He received the Charles L. Mayer
Nature of Light Award in 1949 on the basis of the partly completed manuscripts
 of his `Quantum
Electrodynamics' papers.  The first Einstein prize was awarded to him, along 
with Kurt G\"odel,
in 1951.  The National Medal of Science was presented to him by President 
Johnson in 1964.  The following year, Schwinger, Tomonaga, and Feynman
received the Nobel Prize in Physics from the King of Sweden.

But by this point his extraordinary command of the machinery of quantum field 
theory had convinced
him that it was too elaborate to describe the real world, at least directly.  
In his Nobel Lecture,
he appealed for a phenomenological field theory that would describe directly 
the particles experiencing
the strong interaction.  Within a year, he developed such a theory, Source 
Theory.

\section*{Source Theory and UCLA}

It surely was the difficulty of incorporating strong interactions into
field theory that led to `Particles and Sources,' received
by the {\it Physical Review\/} barely six months after his Nobel lecture, 
in July 1966, based on lectures Schwinger
gave in Tokyo that summer.  One must appreciate the milieu in which 
Schwinger worked in 1966.  For more than
a decade he and his students had been nearly the only exponents of field
theory, as the community sought to understand weak and strong interactions,
and the proliferation of `elementary particles,' through dispersion relations,
Regge poles, current algebra, and the like, most ambitiously through the
$S$-matrix bootstrap hypothesis of Geoffrey Chew and 
Stanley Mandelstam \cite{chew,frautschi,adler,mandelstam}.
 What work in field theory did exist then was largely axiomatic, 
an attempt to turn the structure of the theory into a branch of mathematics,
starting with  Arthur Wightman \cite{wightman},
and carried on by many others, including
Arthur Jaffe at Harvard \cite{jaffe}.
(The name changed from axiomatic field theory to constructive field theory 
along the way.) Schwinger looked
on all of this with considerable distaste; not that he did not appreciate
many of the contributions these techniques offered in specific contexts,
but he could not see how they could form the {\em basis\/} of a theory.

The new source theory was supposed to supersede field theory, much 
as Schwinger's 
successive covariant formulations of quantum electrodynamics had replaced
his earlier schemes.  In fact, the revolution was to be more profound,
because there were no divergences, and no renormalization.
`The concept of renormalization is simply foreign to this phenomenological
theory.  In source theory, we begin by hypothesis with the description of
the actual particles, while renormalization is a field theory concept
in which you begin with the more fundamental operators, which are then
modified by dynamics.  I emphasize that there never can be divergences
in a phenomenological theory.  What one means by that is that one is
recognizing that all further phenomena are consequences of one
phenomenological constant, namely the basic charge unit, which describes the 
probability of emitting a photon relative to the emission of an electron.
When one says that there are no divergences one means that it is not
necessary to introduce any new phenomenological constant.  All further
processes as computed in terms of this primitive interaction automatically
emerge to be finite, and in agreement with those which historically had
evolved much earlier.' \cite{btts}

Robert Finkelstein has offered a perceptive discussion of Schwinger's source
theory program:
`In comparing operator field theory with source theory Julian revealed his 
political orientation when he described operator field theory as a trickle
down theory (after a failed economic theory)---since it descends from implicit
assumptions about unknown phenomena at inaccessible and very high energies to 
make predictions at lower energies.  Source theory on the other hand he
described as anabatic (as in Xenophon's Anabasis) by which he meant that it
began with solid knowledge about known phenomena at accessible energies to make
predictions about physical phenomena at higher energies. Although source theory
was new, it did not represent a complete break with the past but rather was a
natural evolution of Julian's work with operator Green's functions.  His
trilogy on source theory is not only a stunning display of Julian's power
as an analyst but it is also totally in the spirit of the modest scientific
goals he had set in his QED work and which had guided him earlier as a 
nuclear phenomenologist.' \cite{fink}

But the new approach was not well received.  In part this was because
the times were changing; within a few years, 't Hooft \cite{thooft}
 would establish
the renormalizability of the Glashow-Weinberg-Salam $SU(2)\times U(1)$
electroweak model, and field theory was seen by all to be viable again.
With the discovery of asymptotic freedom in 1974 \cite{af},
 a non-Abelian gauge theory of strong interactions,
quantum chromodynamics, which was proposed somewhat earlier \cite{QCD},
was promptly accepted  by nearly
everyone.  An alternative to conventional field theory did not seem to
be required after all. Schwinger's insistence on a clean break with the
past, and his rejection of `rules' as opposed to learning while
serving as an `apprentice,' did not encourage conversions.

Already before the source theory revolution,
 Schwinger felt a growing sense of unease with
his colleagues at Harvard.
But the chief reason Schwinger left Harvard for UCLA was health related.
Formerly overweight and inactive,
he had become health conscious upon the premature death of Wolfgang Pauli
in 1958.  He had been fond of tennis from his youth, had discovered
skiing in 1960, and now his doctor was recommending a daily swim for his
health.  So he listened favorably to the entreaties of David Saxon, 
his closest colleague at the Radiation Lab during the war, who
for years had been trying to induce him to come to UCLA. Very much against
his wife's wishes, he made the move in 1971.  He brought along his three
senior students at the time, Lester DeRaad, Jr., Wu-yang Tsai, and the
present author,  who became long-term `assistants' at UCLA.
He and Saxon expected, as in the early days at Harvard, that
 students would flock
to UCLA to work with him; but they did not.  Schwinger was no longer the
center of theoretical physics.

This is not to say that his little group at UCLA did not make an heroic
attempt to establish a source-theory presence.  Schwinger remained a
gifted innovator and an awesome calculator.  He wrote 2-1/2 volumes of
an exhaustive treatise on source theory, {\it Particles, Sources, and Fields}, 
devoted primarily to the reconstruction of quantum
electrodynamics in the new language; unfortunately, he abandoned the
project when it came time to deal with strong interactions, in part because
he became too busy writing papers on an `anti-parton' interpretation of
the results of deep-inelastic scattering experiments.  
He made some significant contributions
to the theory of magnetic charge; particularly noteworthy was his introduction
of dyons.  He reinvigorated proper-time methods
of calculating processes in strong-field electrodynamics;
and he made some major contributions to the theory of the Casimir effect,
which are still having repercussions.
But it was clear he was reacting, not leading, as witnessed by his
quite pretty paper on the `Multispinor Basis of Fermi-Bose Transformation' 
(1979), in which he kicked himself for not discovering supersymmetry.

\section*{Conclusion}
It is impossible to do justice in a few words 
to the impact of Julian Schwinger on physical thought in the 20th Century.  
He revolutionized fields from nuclear physics to many body theory,
first successfully formulated renormalized quantum electrodynamics,
developed the most powerful functional formulation of quantum field theory,
and proposed new ways of looking at quantum mechanics, angular momentum
theory, and quantum fluctuations.  His legacy includes `theoretical tools'
such as the proper-time method, the quantum action principle, and effective
action techniques.  Not only is he responsible for formulations bearing his
name: the Rarita-Schwinger equation, the Lippmann-Schwinger equation, the
Tomonaga-Schwinger equation, the Schwinger-Dyson equations, 
 the Schwinger mechanism, and so forth, but
some attributed to others, or known anonymously:
Feynman parameters, the Bethe-Salpeter equation,
coherent states, Euclidean field theory; the list goes on and on.
It is impossible to imagine what physics would be like in the 21st Century
without the contributions of Julian Schwinger, a very private
yet wonderful human being.  It is most gratifying that a dozen years after his
death, recognition of his manifold influences is growing, and research projects
he initiated are still underway.

Julian Schwinger lectured twice at the Erice International School on
Subnuclear Physics, in the years 1986 and 1988.


\begin{thebibliography}{99}



\bibitem{MM} Jagdish Mehra and Kimball A. Milton, {\it Climbing the
Mountain: The Scientific Biography of Julian Schwinger\/} (Oxford University
Press, 2000).
\bibitem{QL} K. A. Milton, ed., {\it A Quantum Legacy: Seminal Papers
of Julian Schwinger\/} (World Scientific, Singapore, 2000).

\bibitem{SSS} Silvan S. Schweber, {\it QED and the Men Who Made It: Dyson,
Feynman, Schwinger, and Tomonaga\/} (Princeton University Press, 1994).

\bibitem{erp}A. Einstein,
B. Podolsky, and N. Rosen, {\it Phys.\ Rev.} {\bf 47}, 777 (1935).

\bibitem{quad} J. M. B. Kellogg, I. I. Rabi, N. F. Ramsey, and J. R.
Zacharias, {\it Bull.\ Am.\ Phys.\ Soc.} {\bf 13}, No.~7, Abs.~24 and 15 (1938);
{\it Phys.\ Rev.} {\bf 55}, 318 (1939).

\bibitem{ER} K. A. Milton and J. Schwinger, {\it Electromagnetic
Radiation: Variational Methods, Waveguides, and Accelerators} 
(Springer, Berlin, 2006).

\bibitem{js} Interview with J. Schwinger by Jagdish Mehra, March 1988.

\bibitem{lamb} W. E. Lamb, Jr., and R. C. Retherford, {\it Phys.\ Rev.}
{\bf 72}, 241 (1947).

\bibitem{bethe} H. A. Bethe, {\it Phys.\ Rev.} {\bf 72}, 339 (1947).

\bibitem{g-2} J. E. Nafe, E. B. Nelson,
and I. I. Rabi, {\it Phys.\ Rev.} {\bf 71}, 914 (1947); P. Kusch and H. M.
Foley, {\it Phys.\ Rev.} {\bf 72}, 1256 (1947).

\bibitem{dancoff} S. M. Dancoff, {\it Phys.\ Rev.} {\bf 55}, 959 (1939).

\bibitem{feynman} R. P. Feynman, {\it Phys.\ Rev.} {\bf 76}, 749, 769
(1949).

\bibitem{dyson} F. J. Dyson, {\it Phys.\ Rev.} {\bf 75}, 486, 1736
(1949).

\bibitem{phd} W. Heisenberg and W. Pauli, {\it Zeit.\ f\"ur Phys.}
{\bf 56}, 1 (1929); {\it ibid.} {\bf 59}, 168 (1930);
P.~A.~M.~Dirac, {\it Proc.\ Roy.\ Soc.\
London\/} {\bf A136}, 453 (1932); P. A. M.
Dirac, V. A. Fock, and B. Podolsky, {\it Phys.\ Zeit.\ Sowjetunion\/}
{\bf 2}, 468 (1932).

\bibitem{kramers} H. A. Kramers, {\it Rapports et discussions du 8e Conseil de 
Physique Solvay 1948\/} (Stoop, Bruxelles, 1950), p. 241; M. Dresden,
{\it H. A. Kramers:  Between Transition and Revolution} (Springer-Verlag,
New York, 1987), p.~375;
  H. A. Kramers, {\it Hand- und Jahrbuch der Chemischen Physik I, 
Abschnitt\/} 2 (Leipzig, 1938), p.~89; {\it Nuovo Cim.}~{\bf15}, 108 (1938).

\bibitem{tomonaga} S. Tomonaga, {\it Prog. Theor.\ Phys.} {\bf 1}, 27 (1946);
{\it Phys.\ Rev.} {\bf 74}, 224 (1948).

\bibitem{RTQED} J. Schwinger. `Renormalization Theory of Quantum 
Electrodynamics: An Individual View,' in {\it The Birth of Particle Physics},
ed.~L. M. Brown and L. Hoddeson (Cambridge University Press, 1983), p.~329.

\bibitem{rpf} R. P. Feynman, {\it Phys.\ Rev.} {\bf 74}, 1430 (1948).

\bibitem{fw} J. B. French and V. F. Weisskopf, {\it Phys.\ Rev.}
{\bf 75}, 338 (1949).

\bibitem{kl}  N. M. Kroll and W. E. Lamb, Jr., {\it Phys.\ Rev.} {\bf 75}, 
388 (1949).

\bibitem{lagrangian} P. A. M. Dirac, {\it Phys.\ Zeit.\
Sowjetunion\/} {\bf 3}, 64 (1933).

\bibitem{abj} J. S. Bell and R. Jackiw, {\it Nuovo Cimento\/} {\bf 60A}, 47
(1969); S. L. Adler, {\it Phys.\ Rev.} {\bf 177}, 2426 (1969);
R. Jackiw and K. Johnson, {\it Phys.\ Rev.} {\bf 82}, 1459 (1969). 

\bibitem{brown}  Lowell S. Brown, `An Important Schwinger Legacy:
Theoretical Tools,' talk given at Schwinger Memorial Session at the April 1995
meeting of the APS/AAPT.  Published in {\it Julian
Schwinger: The Physicist, the Teacher, the Man}, ed.\ Y. Jack Ng
(World Scientific, 1996), p.~131.

\bibitem{bs} E. Salpeter and H. Bethe, {\it Phys.\ Rev.} {\bf 84}, 1232 (1951).

\bibitem{gml} M. Gell-Mann and F. Low, {\it Phys.\ Rev.} {\bf 84}, 350 (1951).

\bibitem{adm} R. Arnowitt, S. Deser, and C. W. Misner, {\it Phys.\
Rev.} {\bf 117}, 1595 (1960).

\bibitem{englertbook} J. Schwinger, {\it Quantum Mechanics: Symbolism of
Atomic Measurement}, ed. B.-G. Englert (Springer, Berlin, 2001).

\bibitem{glashow} S. Glashow, {\it Nucl.\ Phys.} {\bf 22}, 579 (1961).

\bibitem{weinberg} S. Weinberg, {\it Phys.\ Rev.\ Lett.} {\bf 19}, 1264 (1967).

\bibitem{chew} For a contemporary account of $S$-matrix theory, see
R. J. Eden, P. V. Landshoff, D. I. Olive, and J. C. Polkinghorne,
{\it The Analytic $S$-Matrix\/} (Cambridge University Press, 1966).

\bibitem{frautschi} For Regge poles, see S. C. Frautschi, {\it 
Regge Poles and S-Matrix Theory\/} (Benjamin, New York, 1963).

\bibitem{adler} For current algebra, see S. L. Adler and R. F. Dashen,
{\it Current Algebras and Applications to Particle Physics\/} (Benjamin,
New York, 1968).

\bibitem{mandelstam} Bootstrap calculations were introduced in
G. F. Chew and S. Mandelstam, {\it Nuovo Cimento\/} {\bf 19}, 752 (1961).
A survey of $S$-matrix theory just before the bootstrap hypothesis may
be found in G. F. Chew, {\it S-Matrix Theory of Strong Interactions\/}
(Benjamin, New York, 1961).

\bibitem{wightman} An accessible early exposition of this approach is
found in R. F. Streater and A. S. Wightman, {\it PCT, Spin and Statistics,
and All That\/} (Benjamin, New York, 1964).

\bibitem{jaffe} For a modern exposition of  some of these ideas, see
J. Glimm and A. Jaffe, {\it Quantum Physics: A Functional Integral Point of
View\/} (Springer-Verlag, New York, 1981).

\bibitem{btts} J. Schwinger, `Back to the Source' in {\it Proceedings of the
1967 International Conference on Particles and Fields}, ed.~C. R. Hagen,
G. Guralnik, and V. A. Mathur (Interscience, New York, 1967), p.~128.

\bibitem{fink} R. Finkelstein, `Julian Schwinger: The QED Period at
Michigan and the Source Theory Period at UCLA' in {\it Julian Schwinger:
The Physicist, the Teacher, and the Man}, ed.\ Y. J. Ng (World Scientific,
Singapore, 1996), p.~105.

\bibitem{thooft} G. 't Hooft, {\it Nucl.\ Phys.} {\bf B33}, 173 (1971);
{\bf B35}, 167 (1971).

\bibitem{af} D. J. Gross and F. Wilczek,
{\it Phys.\ Rev.\ Lett.} {\bf 30}, 1343 (1973);
 H. D. Politzer, {\it Phys.\ Rev.\ Lett.} {\bf 30}, 1346 (1974);
{\it Phys.\ Rep.} {\bf 14C}, 130 (1974).

\bibitem{QCD} M. Gell-Mann, {\it Acta Phys.\ Austriaca Suppl.} 
{\bf IV}, 733 (1972);
H. Fritzsch and M. Gell-Mann, in {\it Proc.\ XVI Int.\ Conf.\ on High
Energy Physics}, ed.\ J. D. Jackson and A. Roberts (National Accelerator
Laboratory, Batavia, IL);
W. A. Bardeen, H. Fritzsch, and M. Gell-Mann, in {\it Scale and Conformal
Symmetry in Hadron Physics}, ed.\ R. Gatto (Wiley, New York, 1973), p.~139.














\end{thebibliography}
\end{document}